\documentclass[12pt]{article}

\usepackage{amssymb,amsfonts,amsmath}
\usepackage{latexsym}
\usepackage[dvips]{graphicx}

\def\i{{\rm i}}
\def\d{{\rm d}}
\def\e{{\rm e}}
\def\del{\partial}
\def\sl{\hspace{-3.4mm}\not\hspace{2.2mm}}

\begin{document}

\title{\large NJL Model at Finite Chemical Potential\\
 in a Constant Magnetic Field
  \footnote{Presented a poster talk to one of the authors at the
   Workshop on {\it Finite Density QCD at Nara}, Japan, July 2003.} 
} 

\author{Tomohiro Inagaki
\and
{\it \normalsize
Information Media Center, Hiroshima University, 
Higashi-Hiroshima,} \\
\and
{\it \normalsize
Hiroshima, 739-8521, Japan} \\
\and
Daiji Kimura and Tsukasa Murata \\
\and
{\it \normalsize
Department of Physics, Hiroshima University, 
Higashi-Hiroshima,} \\
\and
{\it \normalsize
Hiroshima, 739-8526, Japan} 
}

\maketitle

\begin{abstract} 
 We investigate the influence of an external magnetic field
 on chiral symmetry breaking in the Nambu-Jona-Lasinio (NJL) model
 at finite temperature and chemical potential.
 According to the Fock-Schwinger proper-time method, 
 we calculate the effective potential in the leading order of the
 $1/N_{\rm c}$ expansion.
 The phase boundary dividing the symmetric phase and the broken
 phase is illustrated numerically.
 A complex behavior of the phase boundary is found for large 
 chemical potential. 
\end{abstract}

\section{Introduction}

Strong interactions between quarks and gluons are described by
QCD. In resent years much interest has been paid to the phase structure 
of the QCD vacuum. 
A chiral SU($N$)$_{\rm L}\times$SU($N$)$_{\rm R}$
symmetry for quark flavors is broken
down by the QCD dynamics. 
The broken chiral symmetry is restored at high temperature and/or
high density.
One of the most interesting objects may be found in the neutron star 
which is laid near the critical chemical potential at low temperature.
Some of neutron stars have a strong magnetic field.
Therefore we investigate the QCD vacuum state at finite temperature 
and density in a magnetic field.

In the present paper we use the NJL model \cite{Nam} as one of the 
simplest low energy effective theory of QCD and evaluate the effective 
potential in the leading order of $1/N_{\rm c}$ expansion.
The Fock-Schwinger proper-time method \cite{Sch} is applied to a thermal 
field theory to obtain the exact expression of the effective potential
about an external magnetic field.
There are several works to study the NJL model at finite temperature 
and density in an external magnetic field \cite{Chi,Gus}.
Here we use a new formula of the two-point function proposed in 
Ref.~\cite{Ina} where we include the contribution from poles on the 
complex proper-time plane.

The paper is organized as follows.
In the next section we give the explicit expression of the effective 
potential with involving the temperature $T$, chemical potential $\mu$ 
and constant magnetic field $B$ in the NJL model. 
We briefly discuss how to deal with the proper-time integral.
Section 3 is devoted the study of phase structure in the NJL model at
finite $T,\ \mu$ and $B$.
We give some concluding remarks in the last section.

\section{NJL model at finite $T$, $\mu$ and $B$}
NJL model with two flavors is defined by the Lagrangian
\begin{equation}
 {\mathcal L} = \bar{\psi} (\i\del\sl -QA\sl)\psi 
  +\frac{G}{2N_{\rm c}}
  [(\bar{\psi}\psi)^2 +(\bar{\psi} \i\gamma_5\boldsymbol{\tau}\psi)^2] \ ,
 \label{Lag}
\end{equation}
where $N_{\rm c}$ is the number of colors and $G$ is an effective
coupling constant.   
The quark field $\psi$ belongs to a fundamental representation of
the color SU($N$) group and a flavor isodoublet.
$\boldsymbol{\tau}$ represents the isospin Pauli matrices
and $Q$ is the electric charge of the quark fields,
$Q={\rm diag}(2e/3, -e/3)$.
The Lagrangian is invariant
under the global chiral transformation, 
$\psi\to \exp[{\rm i}\theta\gamma_5\tau^3/2]\psi$.
In four space-time dimensions the NJL model is not renormalizable.
We regard the theory as a low energy effective theory stemming from 
a more fundamental theory at a cut-off scale $\Lambda$.

For practical calculations it is more  convenient to introduce 
auxiliary fields $\sigma$ and $\pi$.
The Lagrangian (\ref{Lag}) is rewritten by using the auxiliary fields
\begin{equation}
 {\mathcal L}_{\rm aux} = \bar{\psi} (\i\del\sl -QA\sl)\psi
  -\bar{\psi}(\sigma
   +\i\gamma_5\tau^3\pi)\psi
   -\frac{N_{\rm c}}{2G}(\sigma^2+\pi^2) \ ,
 \label{Lag_aux}
\end{equation}
where $\sigma \simeq -(G/N_{\rm c})\bar{\psi}\psi$ , 
$\pi = \pi^3 \simeq -(G/N_{\rm c})\bar{\psi}\i\gamma_5\tau^3\psi$
and we assume that the ground state dose not break the local
U$(1)_{\rm EM}$ symmetry, i.e. $\langle \pi^\pm\rangle = 0$.
Here we consider a thermal equilibrium state in an external magnetic
field and introduce the temperature in according with the 
imaginary-time formalism.
For a conserved charge of 
$\psi\to \exp[{\rm i}\theta\tau^3/2]\psi$ invariance
we define a quark chemical potential $\mu$ and we assume 
$\mu_u=\mu_d$.  

To find a ground state of the system at finite temperature
and chemical potential in an external magnetic field
we evaluate an effective potential.
In the leading order of the $1/N_{\rm c}$ expansion the effective 
potential $V(\sigma,\pi=0)$ is given by    
\begin{equation}
 V(\sigma,\pi=0) = \frac{1}{4G}\sigma^2 
  - \frac{1}{2\int_0^\beta \d^4x}
  {\rm Tr}\ {\rm \ln}(\i\del\sl +QA\sl -\sigma -\i\gamma_4\mu) \ , 
 \label{V_e1}
\end{equation}
where $\beta=1/T$.
Since the ground state is an isospin singlet, 
we set $\pi=0$.
The ground state of the thermal average of $\sigma$ is given by the
minimum of the effective potential.
If $\sigma$ develops a non-vanishing value of the ground state,
the chiral symmetry is broken down dynamically.

To calculate the effective potential in a strong magnetic field
we use the Fock-Schwinger 
proper-time method.
The second term of the right hand side in  Eq. (\ref{V_e1}) is
rewritten as 
\begin{eqnarray}
  {\rm Tr}\ {\rm ln}(\i\del\sl +QA\sl -\i\mu\gamma_4 -\sigma)
   \simeq -{\rm Tr}\int_0^\sigma \d m\ S(x,x;m)\ , 
  \label{Trln} \\
  (\i\del\sl +QA\sl -\i\mu\gamma_4 -m)S(x,x';m)=\delta^4(x-x')\ .
\end{eqnarray}
In the present paper we consider the constant magnetic field, 
$A_\mu(x)=\delta_{\mu 2}x_1 B$, for simplicity.
The explicit expression of the Green function $S$
is described in Ref.~\cite{Ina}.

\begin{figure}
  \begin{center}
    \resizebox{!}{30mm}{\includegraphics{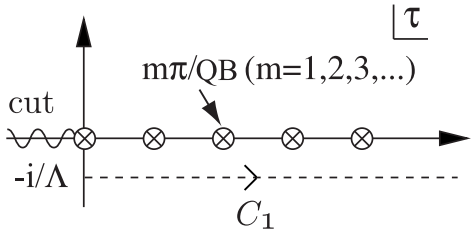}}
    \caption{ The contour of the integration in Eq. (\ref{V_e2}).
       The circles denote the poles.
    }
   \label{path}
  \end{center}
\end{figure}

Substituting the Green function $S(x,x;m)$ to Eq. (\ref{Trln})
and performing the integration about $m$, we obtain the 
second term of the right hand side in Eq. (\ref{V_e1}).
Therefore the effective potential reads
\begin{eqnarray}
 V(\sigma) &=& \frac{1}{4G}\sigma^2
  +\left\{\frac{\e^{-3\i\pi/4}}{4\pi^{3/2}\beta}\sum_{n=0}^{\infty}
  \int_{C_1}\d\tau \right. \nonumber\\
 &&\left.\times\frac{QB\tau}{\tau^{5/2}}\cot(QB\tau)
  \e^{-\i(\omega_n-\i\mu)^2\tau}
  (\e^{-\i\sigma^2\tau}-1)+({\rm c.c.})\right\} \ .
 \label{V_e2}
\end{eqnarray}
It should be noted that we must pay attention to a contour of the 
proper-time integral.
The physical contour is shown as a dashed line in Fig. \ref{path}.

\section{Phase structure}

We perform the summation and the integration in Eq. (\ref{V_e2})
numerically.
The coupling constant $G$ and the proper-time cut-off 
$\Lambda$ are determined to reproduce a pion decay constant (93MeV)
and a pion mass (138MeV).
The value of $G$ and $\Lambda$ depends on the regularization procedure.
In the proper-time cut-off regularization the parameters $G$ and 
$\Lambda$ are determined to be $G=38.7$GeV$^{-2}$ and 
$\Lambda=0.864$GeV with accounting for the current quark mass,  
$m_u+m_d = 14$MeV, see Ref.~\cite{Ina:2}.
We neglect the contribution from the current quark mass to the effective
potential.
Evaluating the minimum of the effective potential we find
the critical value of $T, \mu$ and $B$.

\begin{figure}
  \begin{center}
   \begin{tabular}{cc}
    \resizebox{!}{45mm}{\includegraphics{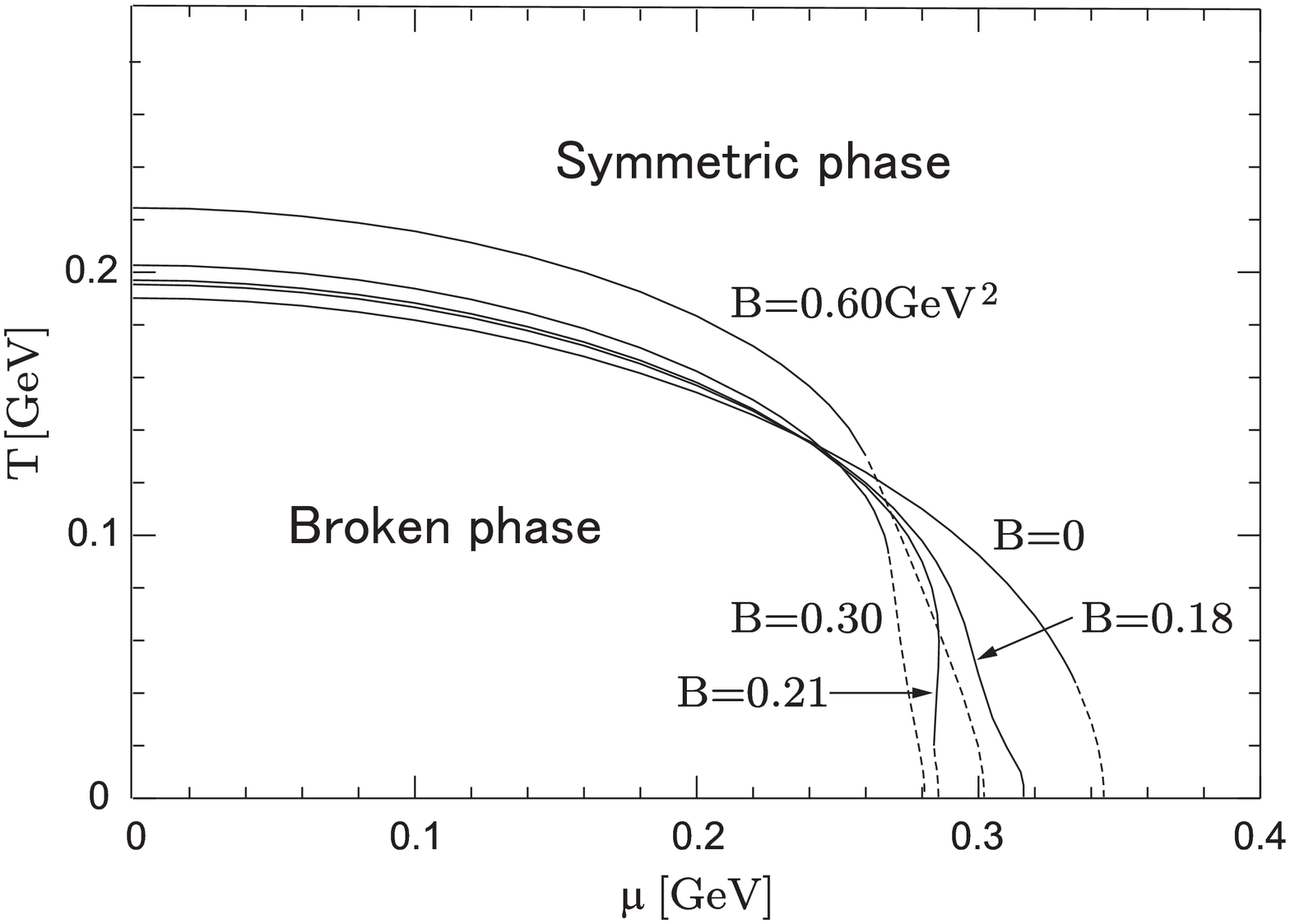}}
   &
    \hspace{3mm}
    \resizebox{!}{45mm}{\includegraphics{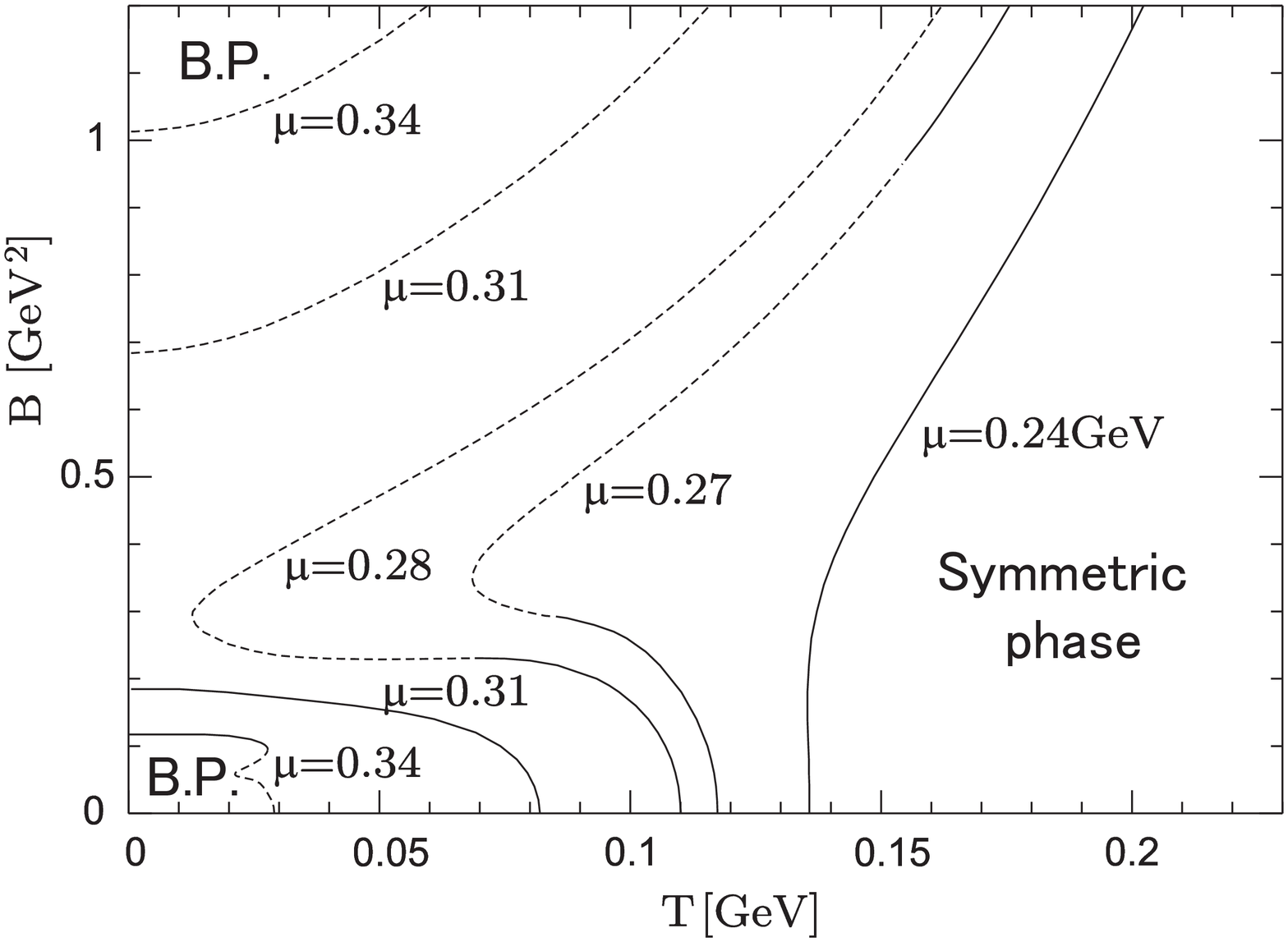}} \\
    {(a) $T-\mu$ plot with $B$ fixed.} 
   &
    {(b) $B-T$ plot with $\mu$ fixed.} 
   \end{tabular}
    \caption{
      Critical lines on $T-\mu$ plane and $B-T$ plane.
      The dashed line represents the first order phase transition 
      while the solid line represents the second  order phase transition.
    }
   \label{plane}
  \end{center}
\end{figure}

Figure \ref{plane}.(a) shows critical lines on the $T-\mu$ plane with 
$B$ fixed.
At $B=0$ the critical $T$ and $\mu$ are a little bit higher than
a known value in a momentum cut-off regularization,
e.g. $T_{\rm cr}|_{\mu=0}\simeq 0.17$GeV \cite{Alfo}.
For small $\mu$ a broken phase spreads out uniformly as $B$ increases.
For large $\mu$ more complex situation is observed.
The broken phase reduces one time near $B\sim 0.2$GeV$^2$.
As is seen in Fig. \ref{plane}.(b) the broken phase is separated into
two parts for $\mu > 0.28$GeV.  
A distortion of critical curve at $\mu=0.34$ GeV is caused by the
de Haas-van Alphen effect \cite{Eber,Shar}.   
From these figures we see that the magnetic field enhances the dynamical 
symmetry breaking below a certain chemical potential, this phenomenon 
is known as the magnetic catalysis \cite{Klim,Suga}.
For large $\mu$ the phase boundary has a different nature in comparison
with the  magnetic catalysis.

\section{Summary}

We have investigated the phase structure of NJL model at finite 
temperature and chemical potential in an external magnetic field.
We calculated the effective potential in the leading order of the 
$1/N_{\rm c}$ expansion by using the Fock-Schwinger proper-time method
generalized to the thermal field theory.
We deal with the combined effects of the temperature, chemical potential
and magnetic field exactly.
We found the phase boundary dividing the symmetric phase and broken 
phase.
For small chemical potential the magnetic catalysis is observed.
The phase boundary shows more a complex structure for large chemical
potential.
It may affect the physics in a dense quark matter.
We will continue our work further and apply our result to some of dense 
objects.

\section*{Acknowledgements}
The authors would like to thank H. Shimizu, M. Fukunaga and 
T. Fujihara for useful conversations.


\begin{thebibliography}{99}
 \bibitem{Nam}
   Y.~Nambu and G.~Jona-Lasinio, 
	Phys. Rev. {\bf 124}, (1961), 246.
 \bibitem{Sch}
   J.~Schwinger, 
        Phys. Rev. {\bf 82}, (1951), 664, \\
   C.~Itzykson and J.~B.~Zuber, 
	{\it Quantum Field Theory} McGrow-Hill Inc. Press, (1980).
 \bibitem{Chi}
   H.~-Y.~ Chiu and V.~Canuto
	Phys. Rev. Lett. {\bf 21}, (1968), 110, \\
   H.~J.~Lee, V.~Canuto, H.~-Y.~Chiu and C.~Chiuderi, 
	Phys. Rev. Lett. {\bf 23}, (1969), 390, \\
   P.~Elmfors, D.~ Persson and B.~-S.~Skagerstam,
	Phys. Rev. Lett. {\bf 71}, (1993), 480.
 \bibitem{Gus}
   V.~P.~Gusynin, V.~A.~Miransky and I.~A.~Shovkovy,
	Phys. Rev. {\bf D52}, (1995), 4718,\\
   S.~Kanemura, H.~T.~Sato and H.~Tochimura,
	Nucl.~Phys.\ {\bf B517}, (1998), 567.
 \bibitem{Ina}
   T.~Inagaki, D.~Kimura and T.~Murata,
	hep-ph/0307289.
 \bibitem{Ina:2}
   T.~Inagaki, D.~Kimura and T.~Murata,
        Prog. Theo. Phys. {\bf 111}, (2004), 371, [hep-ph/0312005].
 \bibitem{Alfo}
   M. Alford, in {\it Dynamics of Gauge Fields: TMU - Yale Symposium},
   edited by A. Chodos, N. Kitazawa, H. Minakata and C. M. Sommerfield,
   Universal Academy Press (2000), [hep-ph/0003185].   
 \bibitem{Eber}
   D.~Ebert and A.~S.~Vshivtsev, 
	hep-ph/9806421, \\
   D.~Ebert, K.~G.~Klimenko, M.~A.~Vdovichenko and A.~S.~Vshivtsev, 
	Phys. Rev. {\bf D61}, (2000), 025005.
 \bibitem{Shar}
   S. G. Sharapov, V. P. Gusynin and H. Beck,
      Phys. Rev. {\bf B69}, (2004), 075104, [cond-mat/0308216].
 \bibitem{Klim}
   K. G. Klimenko, Teor. Mat. Fiz. {\bf 89}, (1991), 211
   and Z. Phys. {\bf C54}, (1992), 323.
 \bibitem{Suga}
   H.~Suganuma and T.~Tatsumi,
	Annals. Phys. {\bf 208}, (1991), 470, \\
   V.~P.~Gusynin, V.~A.~Miransky and I. A. Shovkovy,
	Phys. Rev. Lett. {\bf 73}, (1994), 3499.
\end{thebibliography}
\end{document}